\documentclass[preprint,aps,pre,epsf, superscriptaddress]{revtex4}
\usepackage{amsmath}
\usepackage{amssymb}
\usepackage{epsfig}
\usepackage{graphicx}
\usepackage{multirow}
\usepackage[T1]{fontenc}
\usepackage[utf8]{inputenc}
\usepackage{color}
\usepackage[normalem]{ulem}
\usepackage{tikz}
\usetikzlibrary{positioning}
\usepackage{tabularx}

\begin{document}

\begin{center}
{\Large {\bf Effects of Disorder in a Single-site anisotropic XY ferromagnet: A Monte Carlo study}}\end{center}

\vskip 1cm

\begin{center}{\it Olivia Mallick*}\\
{\it Department of Physics, Presidency University,}\\
{\it 86/1 College Street, Kolkata 700073, INDIA} \\ 
{\it E-mail:mallickolivia0@gmail.com}\\
\end{center}

\vskip 1cm

\noindent {\bf Abstract:} We perform Monte Carlo simulation to study the effects of random disorder on equilibrium phase transition of  three-dimensional single-site anisotropic XY ferromagnet. The disorder is incorporated in two ways; having a randomly distributed anisotropy and presence of a quenched random field. {\color{blue} The ferro-para transition temperature has been found to increase with the increase of the strength of constant (over the space) anisotropy}. In contrast, the system gets ordered at lower temperatures if the anisotropy has random distribution. The effects of quenched random fields are also studied in single-site anisotropic XY ferromagnet. The transition temperature reduces due to the presence of quenched random field. The compensating field (the required amount of field which preserves the critical temperature for isotropic XY ferromagnet) linearly depends on the strength of constant single-site anisotropy. We compute the magnetic and susceptibility exponent ratios for constant single-site anisotropic XY system only via detailed finite-size scaling analysis.  
\vskip 2cm


\vskip 1cm

\noindent
{{\it{*{Present address: Physics of Complex Systems,}\\{\it{ S. N. Bose National Centre for Basic Sciences,}}\\
{\it{JD Block, Sector-III, Salt Lake, Kolkata 700106, India.}}}}

\newpage

\noindent {\bf {\Large 1. Introduction}}\\
Recently, the study of continuous symmetric classical spin systems has drawn plenty of interest. In particular, the XY model has received significant attention owing to explain several physical phenomena. The two-dimensional classical XY model exhibits an unusual kind of phase transition (Kosterlitz-Thouless) without any long-range ferromagnetic order \cite{kosterlitz}. This model is capable of describing some properties of liquid helium, superconductivity, liquid crystal with planner Hamiltonian. The BKT transition was experimentally observed in ultra-cold fermi gas \cite{murthy}.

The critical behaviours of isotropic (SO(2) symmetric) XY ferromagnet have been studied by exact high-temperature series expansion method \cite{betts} and predicted the critical temperature $\frac{kT_c}{J}=4.84\pm0.06$ and estimated the critical exponents. The three dimensional XY ferromagnet has been
investigated \cite{campostrini},\cite{hasenbusch} later by Monte Carlo simulation to estimate the critical temperature (Tc =
2.206...) and specified the XY universality class.
Magnetic anisotropy plays a pivotal role in modifying the physical properties and critical behavior of XY ferromagnets. By destroying the inherent SO(2) symmetry of the isotropic XY model, anisotropy influences vortex dynamics, alters phase transitions, and affects long-range magnetic ordering.  The three-dimensional  XY ferromagnet in presence of {\it bilinear exchange} anisotropy was studied by extensive Monte Carlo simulation via the Metropolis algorithm. This study explores both constant and random anisotropy (with specific statistical distributions) and reports on the phase boundaries and corresponding critical exponents \cite{oliviarandomanis}. The random anisotropy introduces disorder into the system and reduces critical temperature. In this paper, we report findings on the critical behaviour in classical XY model with the inclusion of {\it single-site} anisotropy. The single-site anisotropy also disrupts the continuous SO(2) symmetry but focuses on the extra energy contribution at individual spin sites. The randomly quenched nonmagnetic impurity also plays a role of disorder in the
system. A recent Monte Carlo study \cite{oliviaimpurity} on three-dimensional anisotropic XY ferromagnet shows
that the critical temperature increases linearly with the strength of anisotropy (both bilinear
exchange type and single site type). It should be mentioned that in finite temperature quantum field theoretic calculations, the nonlinear rise of the critical temperature was previously identified for single-site anisotropy \cite{ma}. {\color{blue}Previous studies using MC simulations have shown that non-magnetic impurities in 2D XY model preserve the universality class but affect quasi-long range order in low-temperature \cite{berche}. They analyzed pair correlation, critical exponent variations and supported weak dilution effects with spin-wave like calculations. The quantum phase transition in two dimensional XY system in presence of non-magnetic impurity was studied via quantum Monte Carlo Stochastic Series Expansion and self consistent Harmonic Approximation \cite{lima1}. The BKT transition temperature $T_{BKT}$ is determined as a function of the nonmagnetic impurity density.  The effect of non-magnetic impurities on the quantum phase transitions in 2D antiferromagnet with single-ion anisotropy has been studied using the Boson theory with magnonic exitation \cite{lima2}.} The XY ferromagnet in the presence of magnetic field has been investigated \cite{gouveat},\cite{rosa}. Since
the random field seriously affects the critical behaviour of Ising ferromagnets, it would be an
interesting question, what would be the effect of random magnetic field on the critical behaviours
of XY ferromagnet. Recently, the domain growth and aging are studied \cite{puri} in random field XY
(RFXY) model. The quasi long-range ordered (QLRO) phase in the limit of low disorder (random
field) has been predicted \cite{fisher},\cite{feldman} in the form of Bragg glass phase in random field XY model. A
similar topological phase transition to a pinned vortex-free phase at a nonzero critical field strength
in three-dimensional random field XY systems has been predicted from numerical studies\cite{fisch}. The
vortex-glass phase is also predicted \cite{garanin} from the numerical studies on three-dimensional random
field XY model at zero temperature \cite{kutay}. The arguments show that the lower critical dimension for the quasi-long-range ordered phase is 3.9 (obtained from a functional renormalization group study). So, the long-range ordering is prohibited in two and three dimensions. The range of lower critical dimensions in RFXY model is supported recently. In the presence of a random magnetic field, it would not be surprising to anticipate the order-disorder transition in the three-dimensional XY model for the missing SO(2)symmetric anisotropic XY ferromagnet. A competitive behaviour of the random field and the bilinear exchange anisotropy has recently been investigated
\cite{oliviarfxy} by Monte Carlo simulation. Two different kinds of random fields are used. The magnitude of the random field is constant but the direction is random (i) within $0$ to $2\pi$ (ii) within an angular window that does not have full circular symmetry. The quenched random magnetic fields play a role of disorder to the system and lead to an order-disorder phase transition in an anisotropic RFXY ferromagnet.

In this study we aim to investigate the critical behaviour of the three-dimensional XY ferromagnet in presence of both single-site anisotropy and quenched random field. Through extensive Monte Carlo simulation, we explore how the random distribution of anisotropy or quenched random field introduces disorder into the system and influences the phase transition and critical temperature separately. In addition to that we seek to provide the compensating field which preserve the critical temperature of isotropic XY ferromagnet by systematically varying the strength of anisotropy and random field. The paper is organized as follows: In Section 2, we describe the model and simulation methods employed in this study. Section 3 presents the results of our simulations, divided into two parts: the effect of anisotropy and the quenched random fields on the phase transition temperature and critical behaviour. Section 4 provides a discussion of our findings in the context of previous research followed by conclusions. 

\vskip 0.3cm

\vskip 1cm

\noindent {\bf {\Large 2. Model and Numerical simulations}}\\
\vskip 0.2 cm

The Hamiltonian of classical XY ferromagnet in the presence of single-site anisotropy reads as,
\begin{equation}
   \mathcal{H}=-J\sum_{\langle ij \rangle}[{S_i}^x{S_j}^x+{S_i}^y{S_j}^y] -D\sum_{i}[{({S_i}^x)}^2-{({S_i}^y)}^2]
\end{equation}
 where ${S_i}^x(=\cos{\theta_i})$ and ${S_i}^y(=\sin{\theta_i})$ represent the components of the two-dimensional spin vector having unit length ($|\Vec{S}|=1$) at i-th lattice site. The angle $\theta$ is measured with respect to the positive X-axis and defines the direction of the spin vector $\Vec{S}$. The spins in each lattice site can be viewed as a two-dimensional rotor capable of pointing in any direction determined by $\theta (0 < \theta < 2\pi)$. The coupling $J>0$ denotes the ferromagnetic interaction strength and $\langle ij \rangle$ indicates the summation over distinct nearest neighbours. $D$ represents the single-site anisotropy strength. This orthorhombic anisotropy destroys the rotational symmetry of the XY ferromagnet. When $D=0$, the system exhibits SO(2) symmetry. If $D\neq 0$ the spins tend to align preferentially along a specific direction, leading to anisotropic behaviour.
 
 Expanding on this description, we now address how the disorder is incorporated into the system. {\color{blue} Disorder plays a crucial role in permuting the nature of phase transitions by introducing competing interactions and local inhomogeneities. In the present system, disorder is introduced in two different ways: \textbf{site-dependent anisotropy} and \textbf{random fields} both of which significantly impact the stability of the ordered phase and transition temperature. Specifically, we consider anisotropy to be site-dependent ($D_i$), characterized by a random statistical distribution throughout the lattice. The disorder is imposed in such a way that the global average over the entire system remains zero i.e, $\langle D_i \rangle =0$.This condition ensures that, on average the system does not exhibit a preferred alignment of anisotropy. However, at the local level, for site-to-site variations in $D_i$, spins struggle to align in a uniform fashion across the lattice, effectively altering the long-range order. Anisotropy is treated as a random variable with three different types of statistical distributions.}

\noindent\textbf{\textit{a) Uniformly Distributed Random Anisotropy:}} The anisotropy variable follows a uniform statistical distribution.

\begin{equation*}
    P_u(D_i)=\frac{1}{w}
\end{equation*}
where $w$ is the width of the distribution. The anisotropy takes the values in the range $-\frac{w}{2}$ to $\frac{w}{2}$. The standard deviation of the distribution is $\sigma_u=\frac{w}{2\sqrt{3}}$.

\noindent\textbf{\textit{b) Normally distributed random anisotropy:}} We choose the site-dependent random anisotropy having the probability distribution of the form,
\begin{equation*}
    P_n(D_i)=\frac{1}{\sqrt{2\pi}w}\exp{(-\frac{{D_i}^2}{2w^2})}
\end{equation*}
 This Gaussian distribution is obtained by the Box-Muller algorithm \cite{box1958note}. The standard deviation of the distribution is $\sigma_n=w$.

 \noindent\textbf{\textit{c) Bimodal distribution of random anisotropy:}} The anisotropy has the probability distribution, 
 \begin{equation*}
     P_b(D_i)=\frac{1}{2}[\delta(D_i -\frac{w}{2})+\delta(D_i +\frac{w}{2})]
 \end{equation*}
where $\delta$ represents the Dirac delta function. The distribution indicates that the values of anisotropy can be either $-w/2$ or $+w/2$ with equal probability. The standard distribution corresponds to Bimodal distribution is given by $\sigma_b=\frac{w}{2}$.

Next, the Hamiltonian of anisotropic XY ferromagnet in presence of random field is given by,
\begin{equation}
   \mathcal{H}=-J\sum_{\langle ij \rangle}[{S_i}^x{S_j}^x+{S_i}^y{S_j}^y] -D\sum_{i}[{({S_i}^x)}^2-{({S_i}^y)}^2]-\sum_{i}\Vec{h_i}\cdot \Vec{S_i}
\end{equation}

The last term in Eq-(2) corresponds to the interaction of individual spins to the random field $\Vec{h_i}(=h\cos{\phi_i},h\sin{\phi_i})$. The randomness of the field is governed by the angle $\phi_i$. The magnitude of the field strength $h$ is constant, where the direction $\phi_i$ can vary uniformly from $0$ to $2\pi$. Thus the average (over the lattice) value of the random field is $\langle \Vec{h_i}\rangle=0$. The magnitude of the field is measured in unit of $J$. {\color{blue}This implies that although individual spins experience a local random field, the system does not have a net external field on average. The random field incorporates competing local energy landscapes, making it more difficult for spins to align coherently. This weakens the ferro-para phase transition temperatures. }

In order to study, critical behaviour and equilibrium phase transition in disordered single-site anisotropic XY ferromagnet we use Monte Carlo simulation with random update of spins via Metropolis algorithm.
 A three-dimensional simple cubic lattice of size L (=20 here, apart from the finite size analysis) is considered here.  We have applied the conventional periodic boundary conditions in all three directions of the lattice. The lattice dimensions of the system are three (simple cubic) and the dimensions of
the spin vector are two (planar ferromagnet or XY model). We can briefly describe the simulation procedure as follows.
The Monte Carlo simulation starts from a random initial configuration of the spin, corresponding to a very high-temperature phase. At a finite temperature T (measured in the unit of $J/k_B$, where $k_B$ is Boltzmann
constant), a lattice site (say x,y,z) is chosen randomly (at any instant of time t) having a random initial spin
configuration (represented by an angle $\theta_i(x, y, z, t)$). A new configuration of the spin (at site x,y,z) is also chosen
(represented by $\theta_f (x, y, z, t')$) randomly. {\color{blue} This represent a trial configuration which the system may accept or reject based on the Metropolis criterion.} The change in energy ($\delta \mathcal{H}$) due to the change in configuration (angle) of
spin (from $\theta_i(x, y, z, t)$ to $\theta_f(x, y, z, t')$) is calculated from Eq-(1) and Eq-(2) for respective cases. The probability
of accepting the new configuration is calculated from the Metropolis formula,
$P_f = Min[\exp{(-\frac{\delta \mathcal{H}}{k_BT})},1]$\cite{binder}. 
{\color{blue} The probabilistic criterion ensures that energetically favorable spin configurations (i.e those reduce the system's energy) are always accepted, while unfavorable configurations (those increase energy) are accepted with a probability that decreases exponentially with $\delta \mathcal{H}$. A uniform random number $r \in [0,1]$ is generated. If $r \leq P_f$, the proposed spin configuration is accepted and the site updates it state accordingly. Otherwise, the site retains its previous configuration. In this way, $L^3$ number of sites
are updated randomly. $L^3$ number of such random updates constitutes on Monte Carlo
step per site (MCSS), which serves as the fundamental unit of time in the simulation process.} Throughout the study the
system size L(= 20). The total length of simulation is $2 \times 10^4$ MCSS, out of which the initial $10^4$ MCSS times
are discarded. This amount of transient steps is sufficient for the system to reach ergodicity. All statistical quantities are
calculated by averaging over the rest $10^4$ MCSS.

 The instantaneous components of magnetisation as follows, 
 \begin{equation*}
     m_x(t)=\frac{1}{L^3}\sum_i {S_i}^x(x,y,z,t)= \frac{1}{L^3}\sum_i \cos{\theta_i(x,y,z,t)}
 \end{equation*}
 and 
 \begin{equation*}
      m_y(t)=\frac{1}{L^3}\sum_i {S_i}^y(x,y,z,t)= \frac{1}{L^3}\sum_i \sin{\theta_i(x,y,z,t)}
 \end{equation*}

The total equilibrium magnetisation is defined as $m=\sqrt{{m_x}^2+{m_y}^2}$. The corresponding susceptibility is determined by, 
 \begin{equation*}
     \chi=\frac{L^3}{k_B T}(\langle m^2 \rangle -\langle m \rangle ^2)
 \end{equation*}
In order to determine the magnetic phase transition in the thermodynamic limit ($L\rightarrow \infty$), we calculate the thermal variation of fourth-order Binder Cumulant for different system sizes which is defined as,
 \begin{equation*}
     U_L=1-\frac{\langle m ^4 \rangle}{3\langle m^2 \rangle ^2}
 \end{equation*}
 The symbol $\langle ..\rangle$ corresponds to time averaging, which is approximately equal to the ensemble averaging in the ergodic limit. The above observable quantities are further averaged over some (20 to 200) different random disorder realizations.

\vskip 0.75cm

\noindent {\bf {\Large 3. Results}}\\

\vskip 0.2cm

In this section, we present our MC simulation results for single-site anisotropic XY ferromagnet in presence of two different disorder environments:(A) Random distribution of anisotropy and (B) Quenched Random field.
\newpage

\noindent {\large\bf {A. Random Distribution of Anisotropy}}\\
\vskip 0.2 cm
Firstly, we discuss the dependence of transition temperature in presence of constant anisotropy. $D$ is considered constant throughout the system. Each lattice site experiences same strength of anisotropy ($D$). Initially, the system is at a high temperature with random configurations of spin. This corresponds to zero magnetisation paramagnetic phase. The system is slowly cooled down with small temperature step $\delta T=0.05$. As the temperature decreases the magnetisation is found to take a non-zero value(for a particular strength of anisotropy, say $D=1.0$), and the system gets ferromagnetically ordered at some transition temperature. Figure-\ref{fig:const-D}(a) presents the thermal variation of magnetisation for two different strengths of anisotropy.  The thermal variation of corresponding susceptibility is shown in  Figure-\ref{fig:const-D}(b). The susceptibility exhibits a peak at the transition point which indicates the second-order phase transition. The transition temperature or critical temperature (for finite-sized system, we may call it pseudocritical temperature) is determined from the position of the peak of susceptibility. From Figure-\ref{fig:const-D}(b) one can observe that the transition point shifts towards high temperature for a higher strength of constant anisotropy.
\begin{figure}[ht!]
    \centering
   (a)\includegraphics[width=0.45\linewidth]{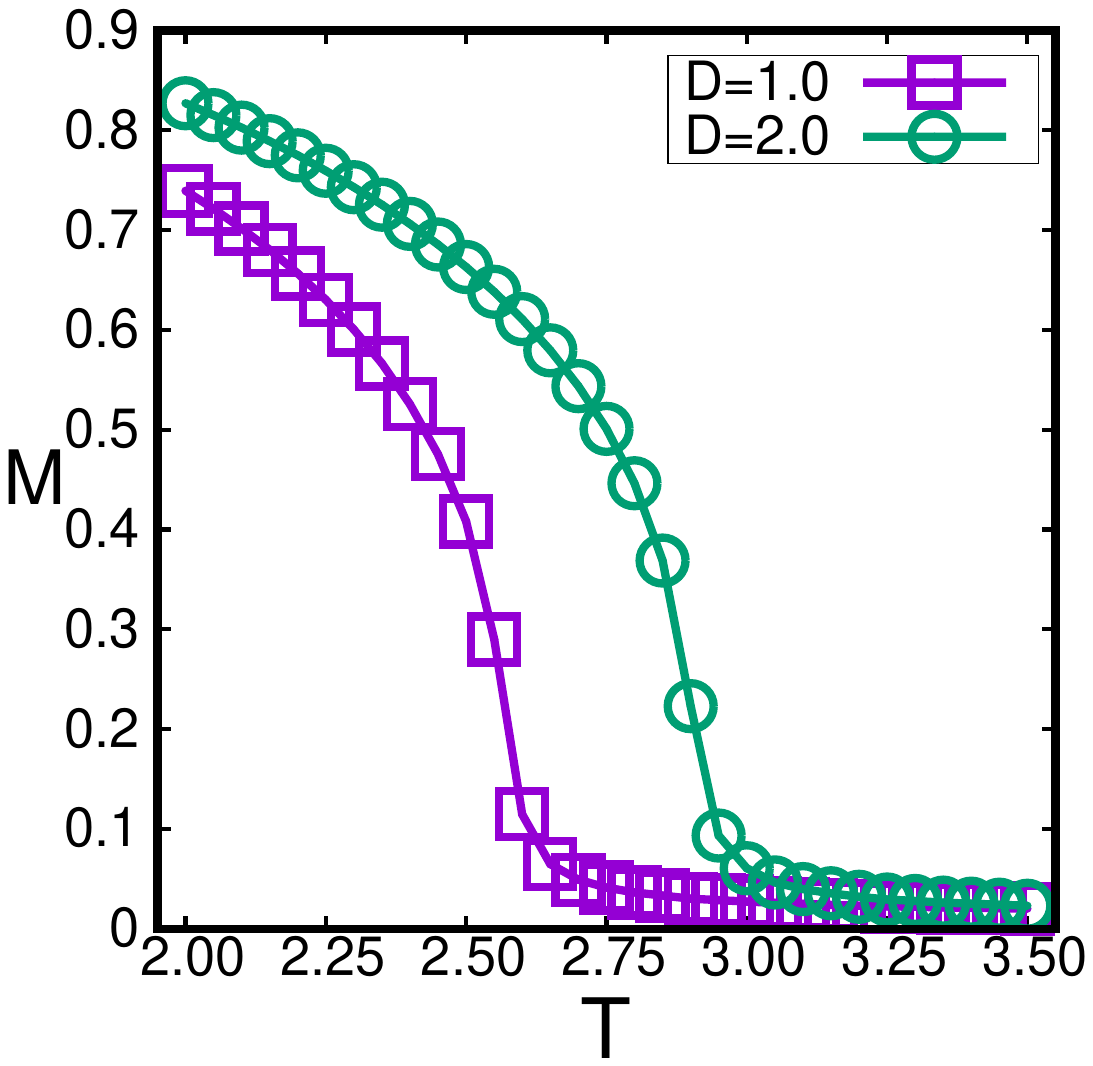}
    (b)\includegraphics[width=0.45\linewidth]{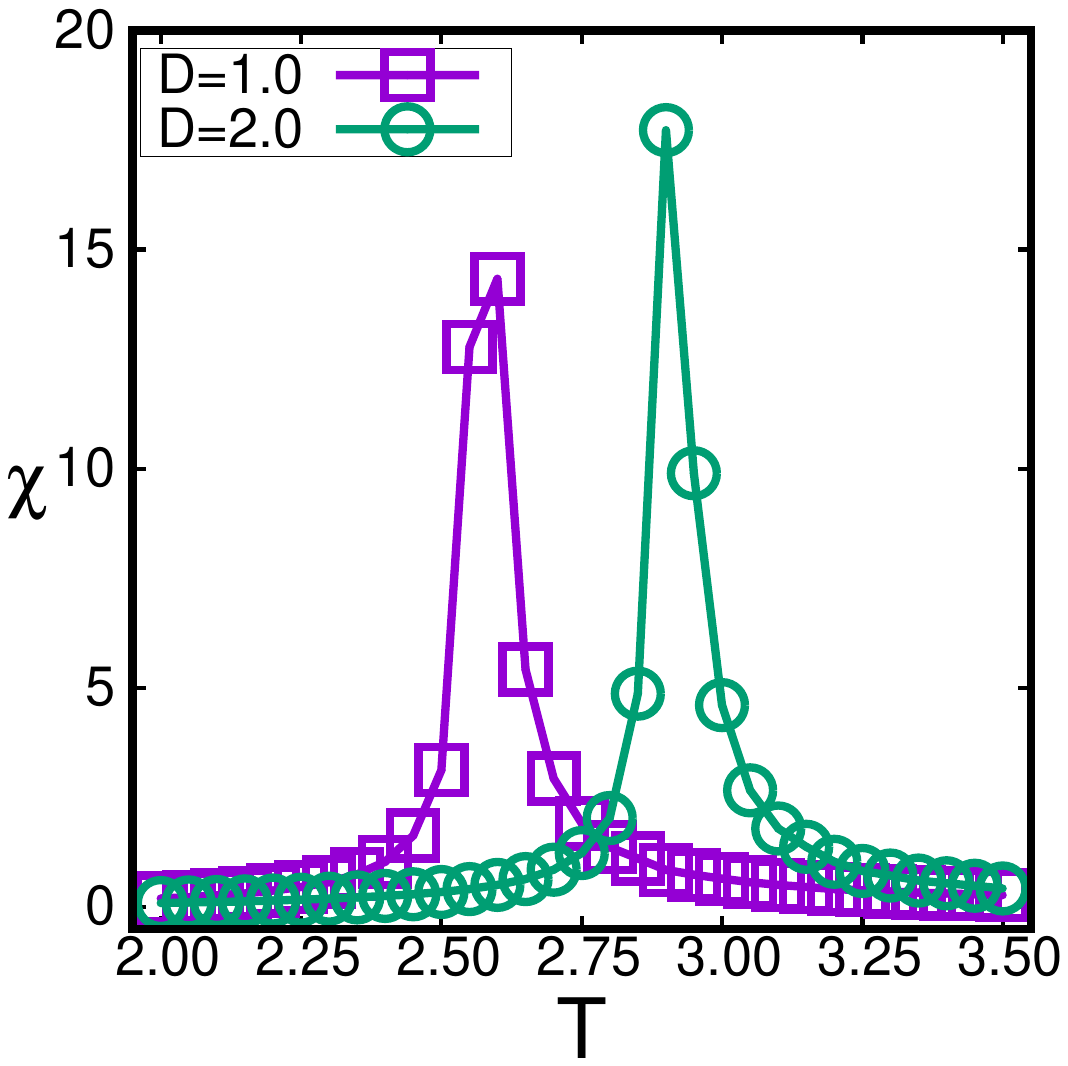}
    \caption{\small{Thermal variation of (a) Magnetisation ($M$) and (b) susceptibility ($\chi$) for two different strength of single-site anisotropy $D=1.0,2.0$.}}
    \label{fig:const-D}
\end{figure}
Figure-\ref{fig:D-const-anis-phase} presents the phase diagram separating the ordered and disordered phases in the $D-{T_c}^*$ plane. This phase diagram reveals a linear relationship between the transition temperature (pseudocritical temperature) and the anisotropy strength. As $D$ increases, ${T_c}^*$ rises proportionally, indicating that the stronger anisotropy stabilizes the ordered phase by enhancing the system's resistance to thermal fluctuations. This linear trend allows us to extrapolate the transition temperature to $D=0$ where the system reverts to the isotropic XY model. By performing a straight line fit, the critical temperature for $D=0$ is estimated as $T_c=2.26 J/k_B$, which aligns close to the known $T_c$ of isotropic XY ferromagnet \cite{campostrini}.\\
\begin{figure}[ht!]
    \centering
    \includegraphics[width=0.65\linewidth]{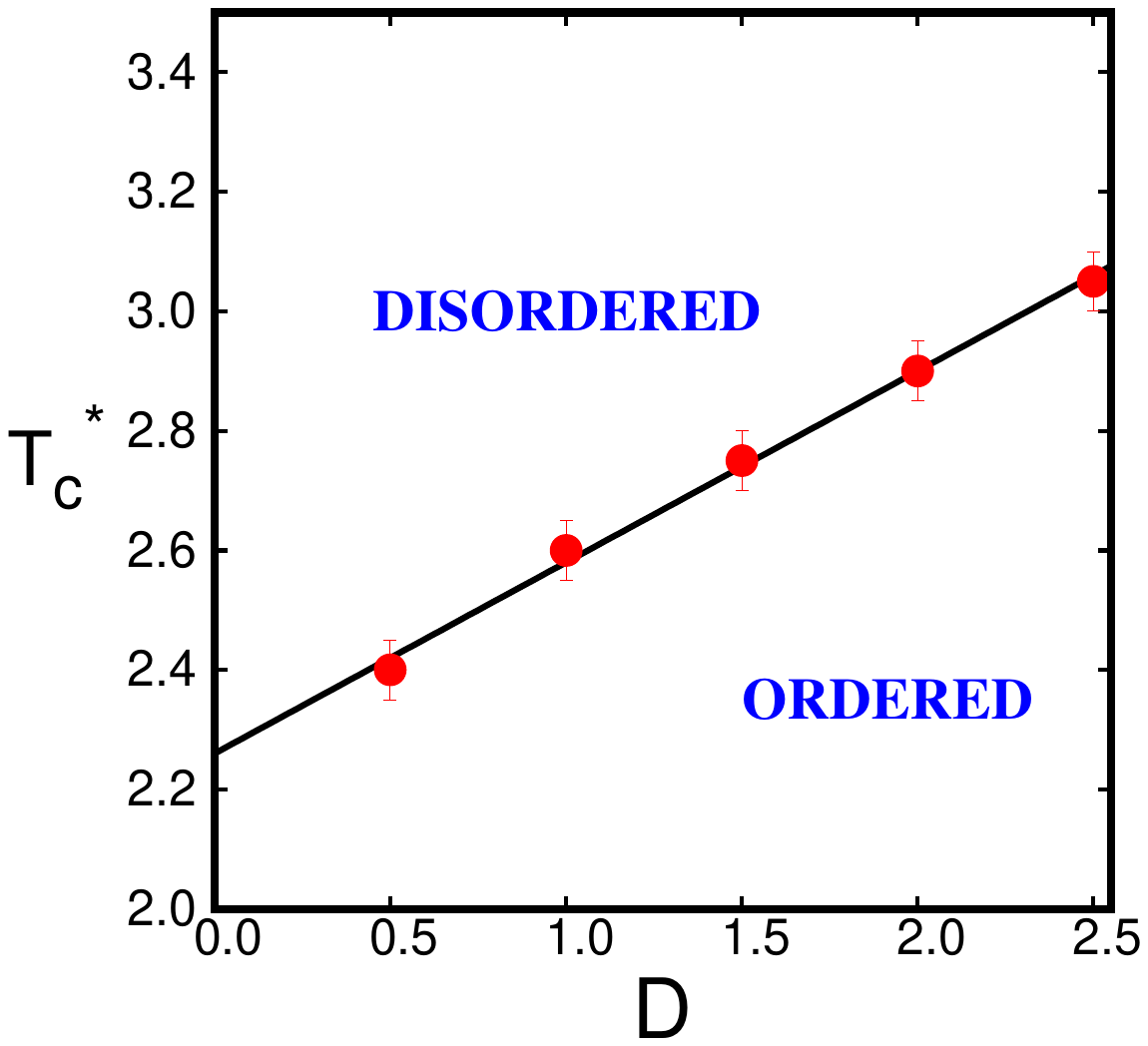}
    \caption{The pseudocritical temperature (${T_c}^*$) as a function of anisotropy D . The fitted straight line equation is ${T_c}^* = aD + b$, with coefficients $a = 0.32\pm 0.01$
 and $b= 2.26\pm 0.02$ . }
    \label{fig:D-const-anis-phase}
\end{figure}
The linear increase in $T_c$ with $D$ underscores the significant role of single-site anisotropy in enhancing the thermal stability of the ordered phase. This behaviour suggests that anisotropy reinforces the alignment of spins along a preferred direction, making it more robust for thermal disturbances to disrupt the ordered state. As a result, a higher temperature is required to prompt the phase transition. The phase diagram observed for single-site anisotropy mirrors the behaviour noted in systems with bilinear exchange anisotropy\cite{oliviarandomanis}. Both types of anisotropy—single-site and bilinear exchange—destroy the SO(2) symmetry of the XY system by favouring preferred spin alignment, leading to similar phase diagram behaviours. The symmetry breaking in presence of anisotropy reinforces the system's bias towards ordered phases, resulting in comparable effects on the transition temperature and critical properties.\\
Next, the anisotropy is considered as a random variable. The effects of randomly distributed single-site anisotropy on the phase behaviour of the system are studied for three distinct types of statistical distributions: Uniform, Gaussian, and Bimodal. Despite the inherent differences in these distributions, where the anisotropy strength ($D_i$) is randomly assigned to each site with an average value of zero across the system ($\langle D_i \rangle=0$), we observe a consistent pattern of results across all three cases.
\begin{figure}[h!]
    \centering
    (a)\includegraphics[width=0.45\linewidth]{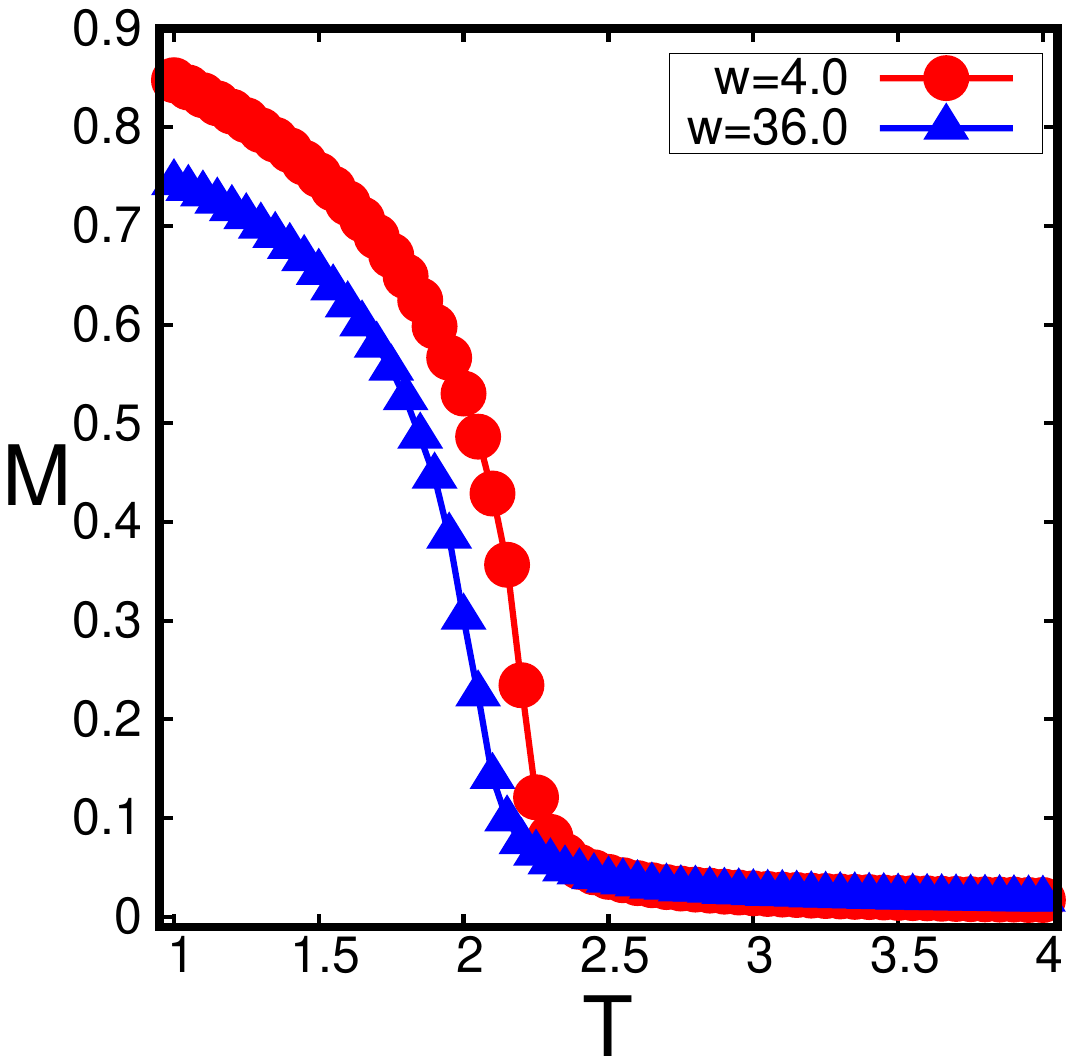}
    (b)\includegraphics[width=0.45\linewidth]{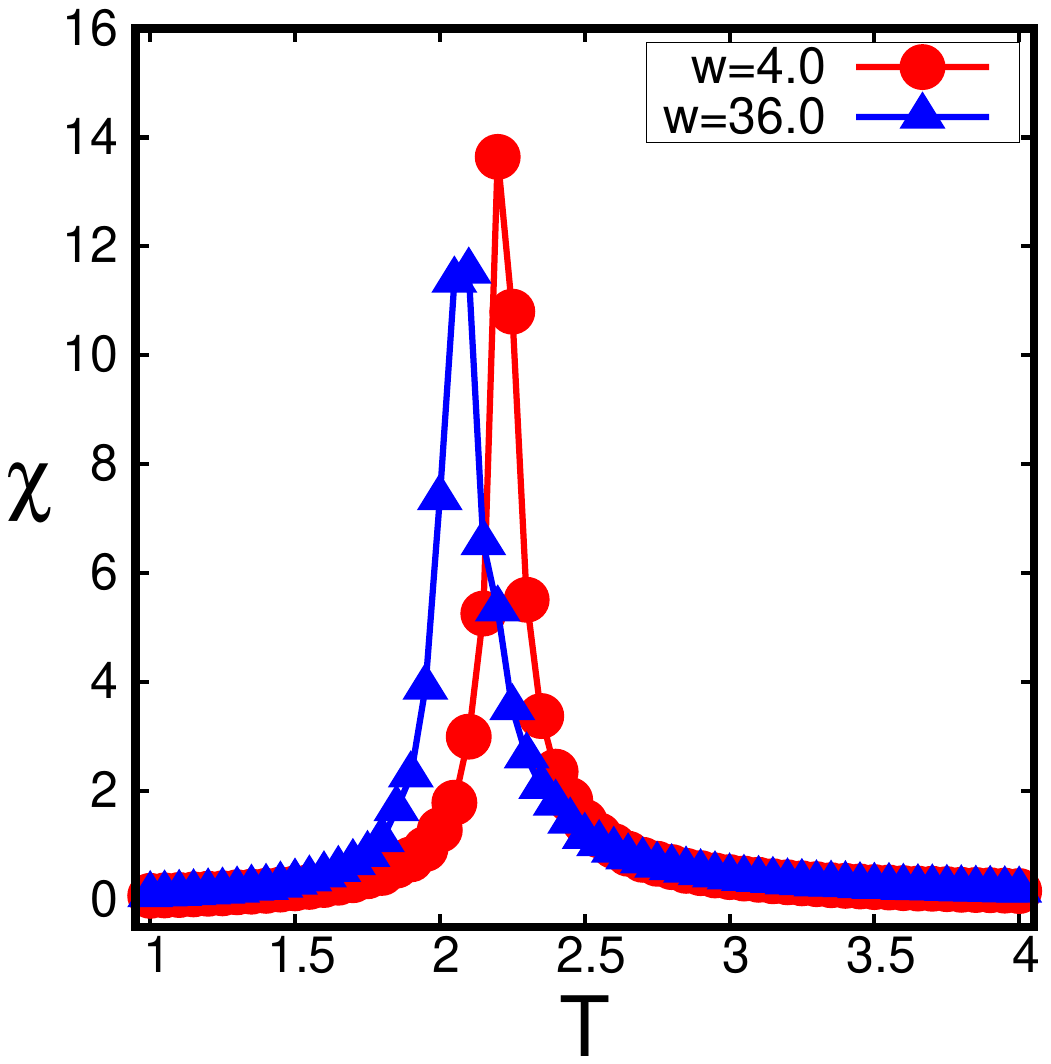}
    (c)\includegraphics[width=0.45\linewidth]{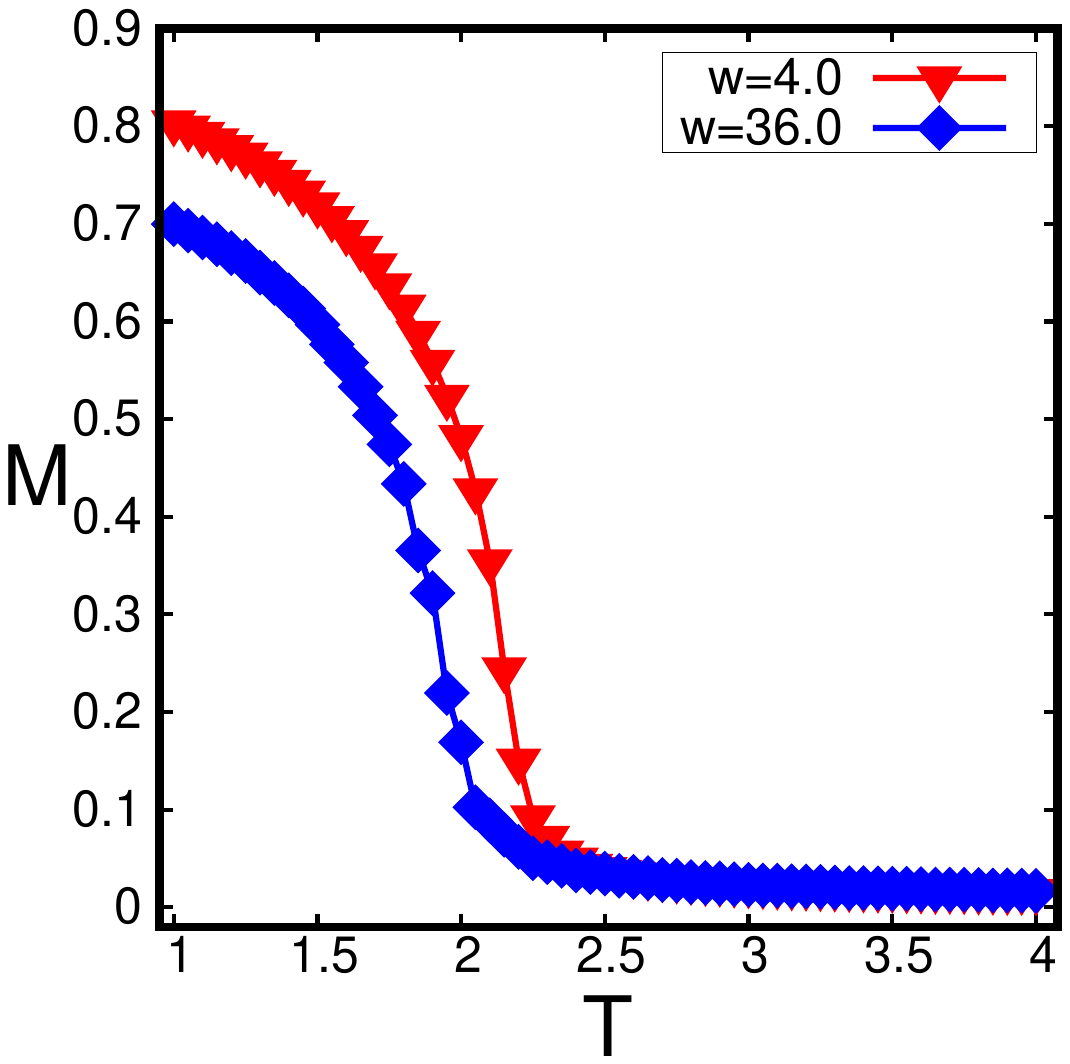}
    (d)\includegraphics[width=0.45\linewidth]{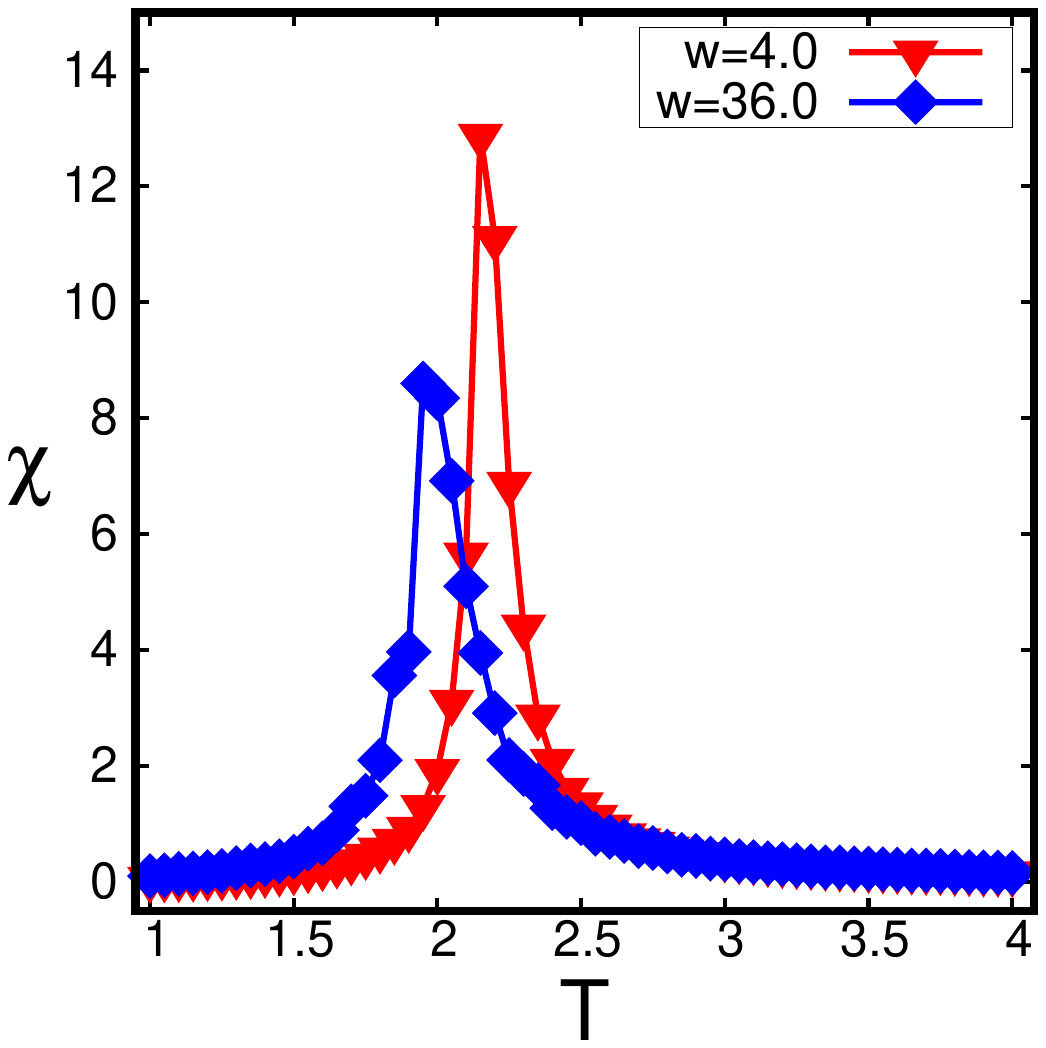}
    (e)\includegraphics[width=0.45\linewidth]{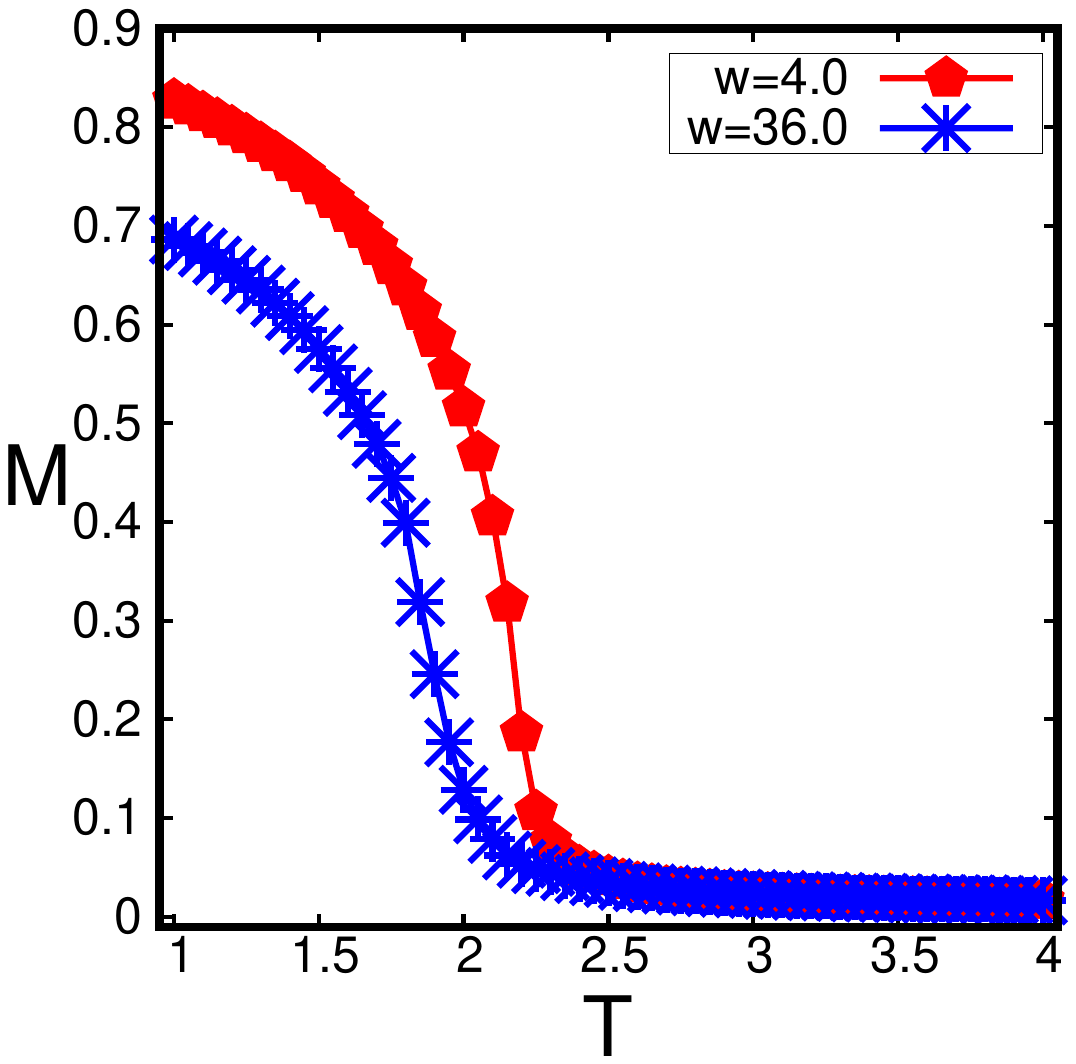}
    (f)\includegraphics[width=0.45\linewidth]{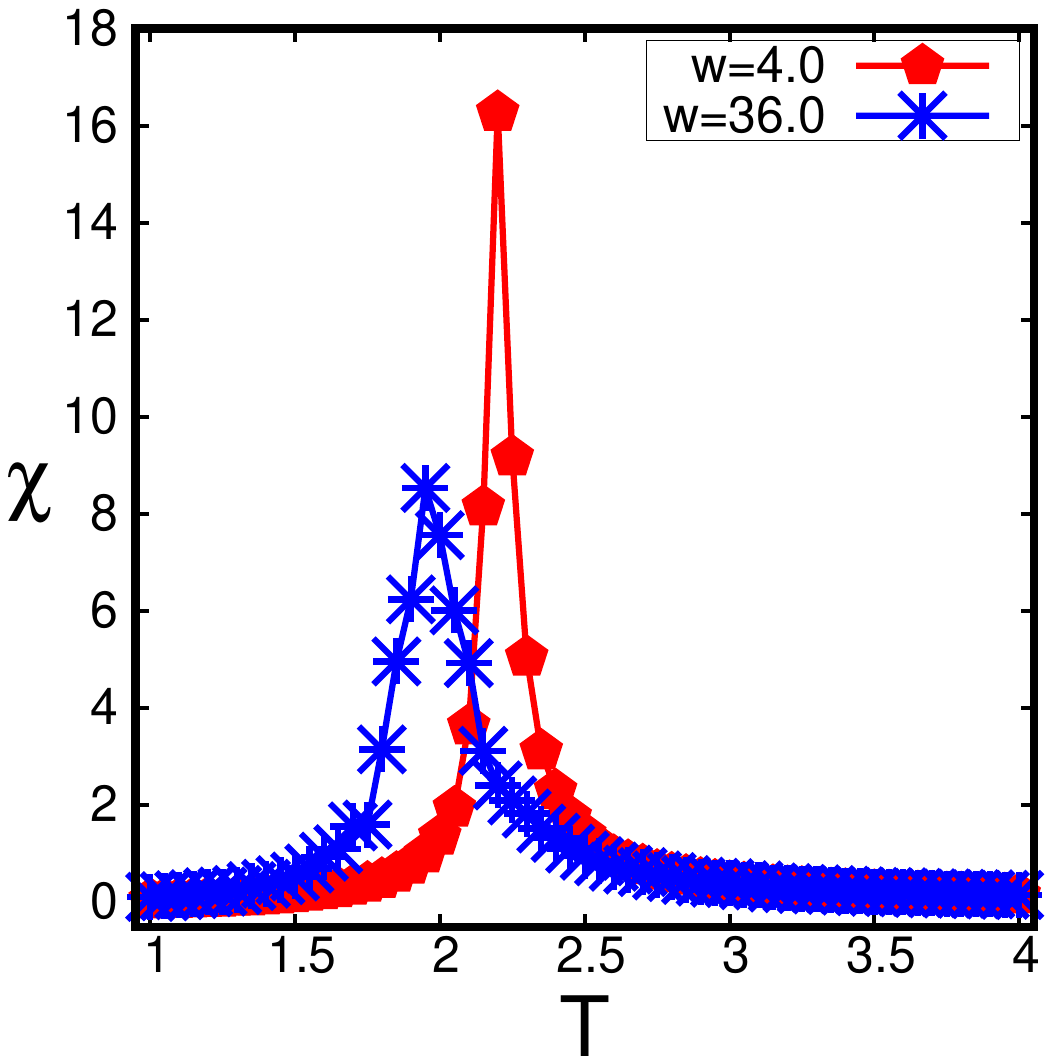}
    \caption{The thermal variation of magnetisation and corresponding susceptibility for three different distributed random anisotropy. The results are shown as Upper panel:(a) $\&$ (b) Uniform distribution, middle panel:(c) $\&$ (d) Gaussian distribution and lower panel:(e) $\&$ (f) Bimodal distribution.}
    \label{fig:D-dist-anis-observables}
\end{figure}

For each type of anisotropy distribution, the system demonstrates a clear transition from a disordered paramagnetic state to an ordered ferromagnetic state as the temperature decreases. This transition is characterized by a steady growth in magnetization, signifying the alignment of spins as thermal fluctuations diminish. The corresponding susceptibility exhibits a sharp peak at a particular temperature. This peak marks the transition temperature (or pseudocritical temperature ${T_c}^*$ for finite-sized system), signifying the point at which the system undergoes a phase transition from the paramagnetic to the ferromagnetic phase. Figure- \ref{fig:D-dist-anis-observables} illustrates the results for three types of distributed anisotropy. One can observe that the maxima of susceptibility curves shift towards lower temperatures as the width of the distribution increases for each type of anisotropy distribution. The critical temperature decreases. The variation of critical temperature with the width of distribution is shown in figure-\ref{fig:D-dist-anis-phase}. However, this decrease in ${T_c}^*$ is relatively modest in the case of single-site anisotropy, indicating that the system's phase transition is less sensitive (in comparison to that for the case of bilinear exchange kind of anisotropy) to variations in the anisotropy distribution width.

\begin{figure}[h!]
    \centering
    \includegraphics[width=0.65\linewidth]{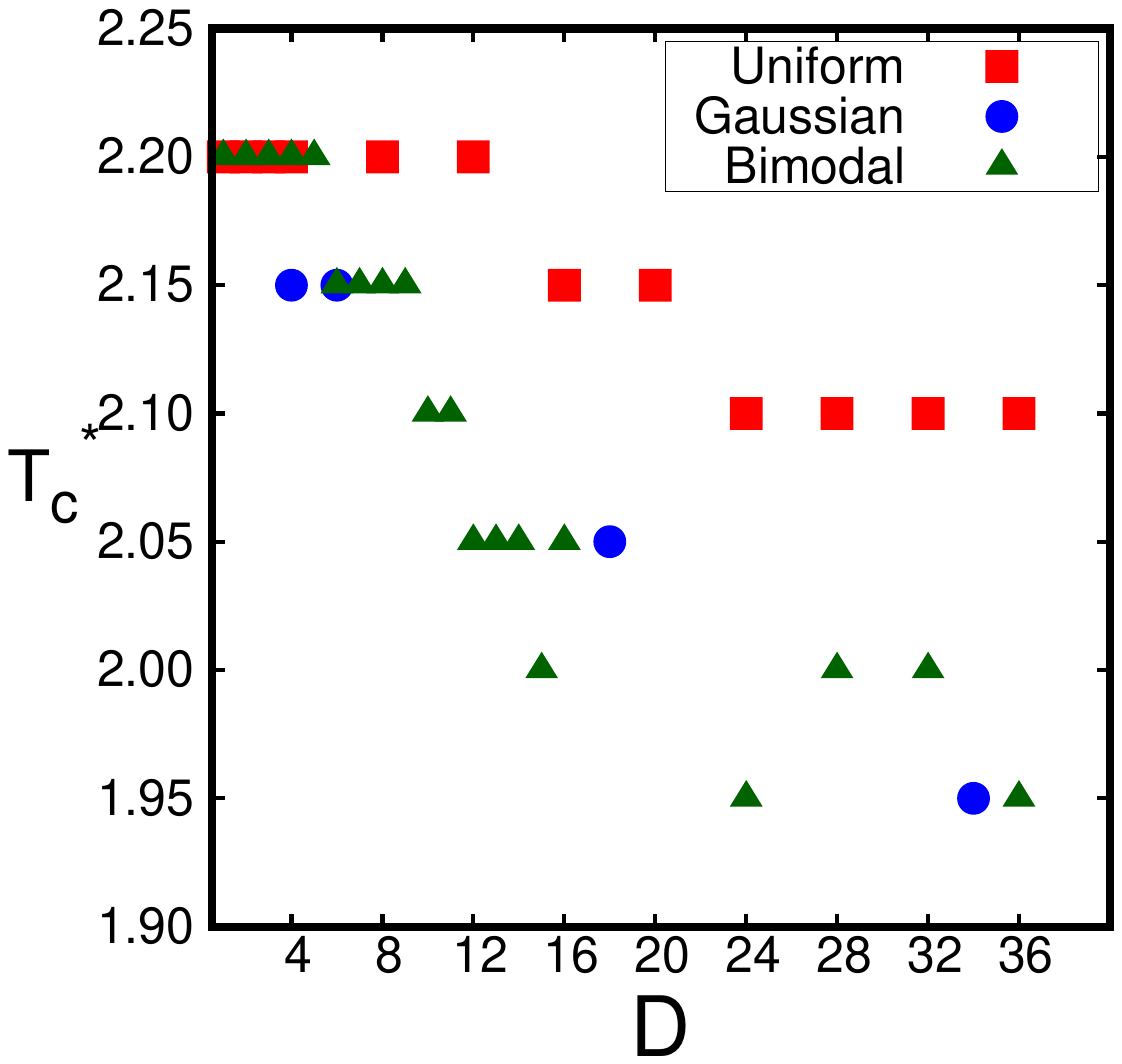}
    \caption{Phase boundary for single-site ($D$) distributed anisotropy. The pseudo-critical temperature(${T_c}^*$) is plotted as a function of width of distribution for three
different kinds of distributed anisotropy.}
    \label{fig:D-dist-anis-phase}
\end{figure}
In contrast, the bilinear exchange anisotropy shows a more pronounced decrease in the transition temperature with increasing distribution width. This difference suggests that bilinear exchange anisotropy has a greater impact on the system's critical behaviour, making the transition temperature more sensitive to the spread of anisotropy values.\\
Finally, the equilibrium phase transition in the three-dimensional XY ferromagnet in presence of single-site anisotropy is confirmed by finite size scaling analysis. The finite size effects have been calculated for constant $D=2.0$. Figure-\ref{fig:const-D-Ul-L} depicts the typical fourth-order Binder cumulant curves as function of temperature for various system sizes $L=10,15,20,25$. The intersection point determines the critical temperature $T_c=3.01$ in thermodynamic limit $L\rightarrow \infty$ for $D=2.0$.
\begin{figure}[ht!]
    \centering
    \includegraphics[width=0.65\linewidth]{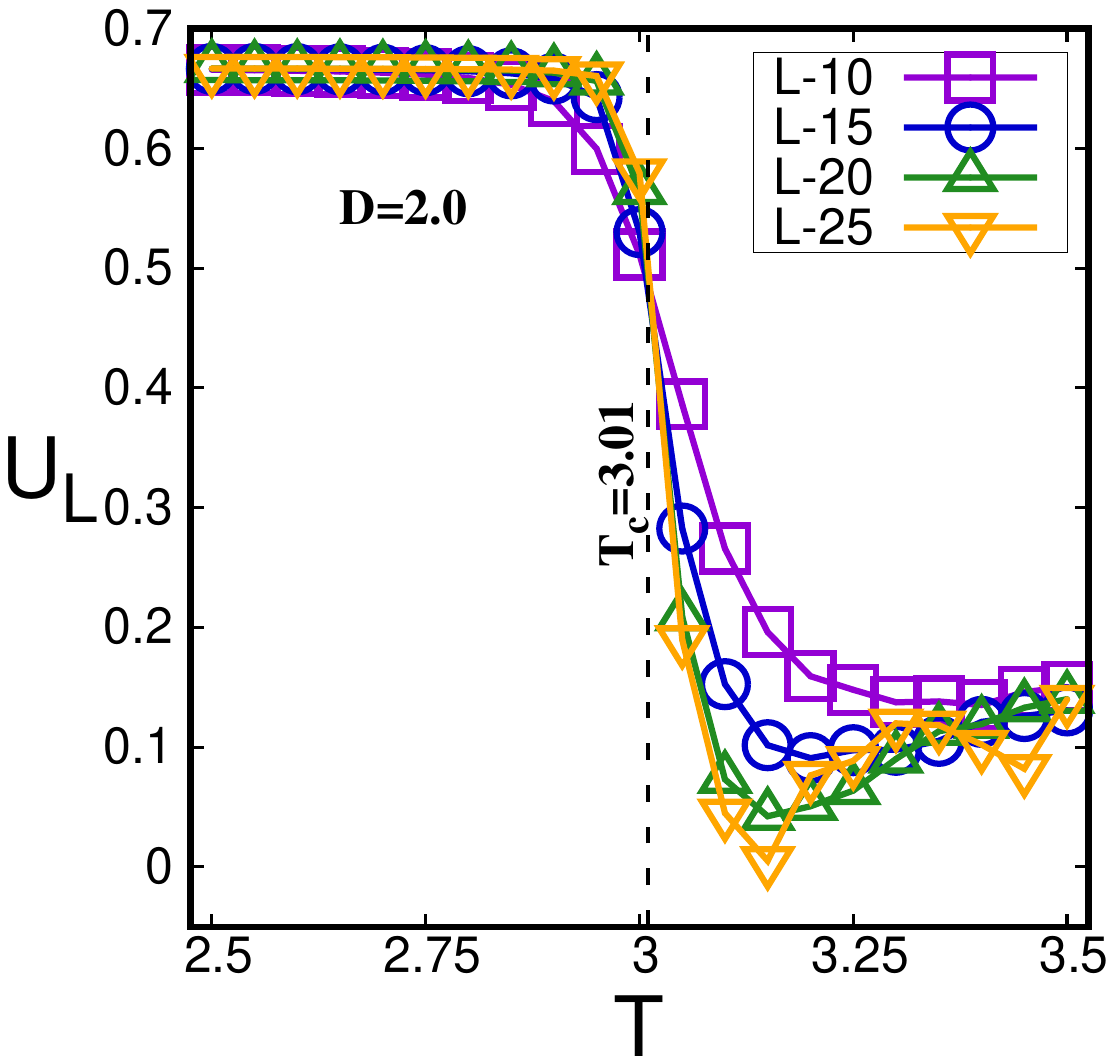}
    \caption{ \small{Fourth order Binder cumulant as a function of temperature for different system sizes at constant strength of anisotropy $D=2.0$. The vertical line passing through the common intersection represents the critical temperature ${T_c} = 3.01J/k_B$.}}
    \label{fig:const-D-Ul-L}
\end{figure}
 The thermal variation of magnetization ($M$) for various system sizes is shown in Figure-\ref{fig:const-D-M-L}(a). The
vertical line ($T_c=3.01$) intersects the curves at different points and determines magnetisations at critical temperature($M(T_c)$).
\begin{figure}[ht!]
    \centering
    (a)\includegraphics[width=0.45\linewidth]{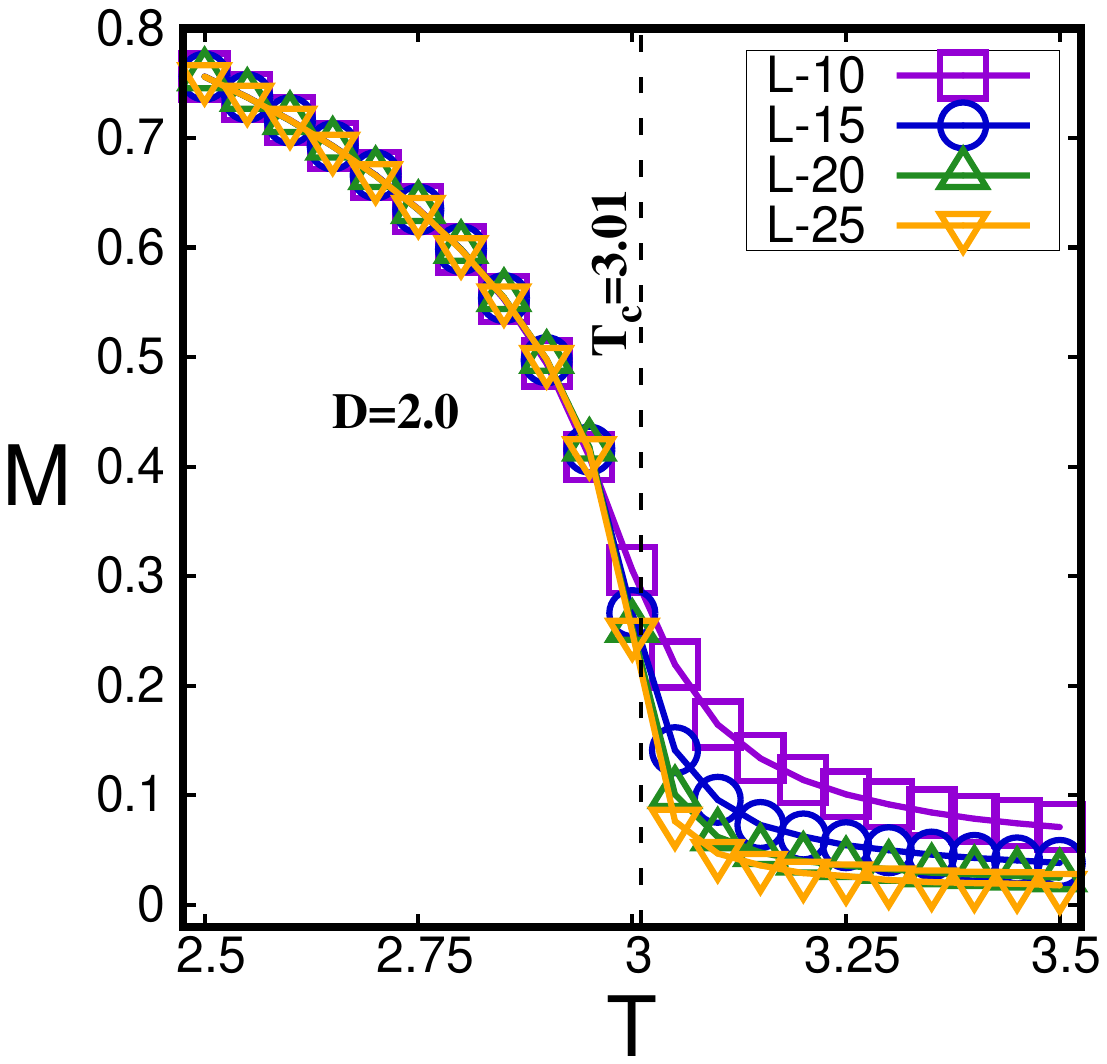}
    (b)\includegraphics[width=0.45\linewidth]{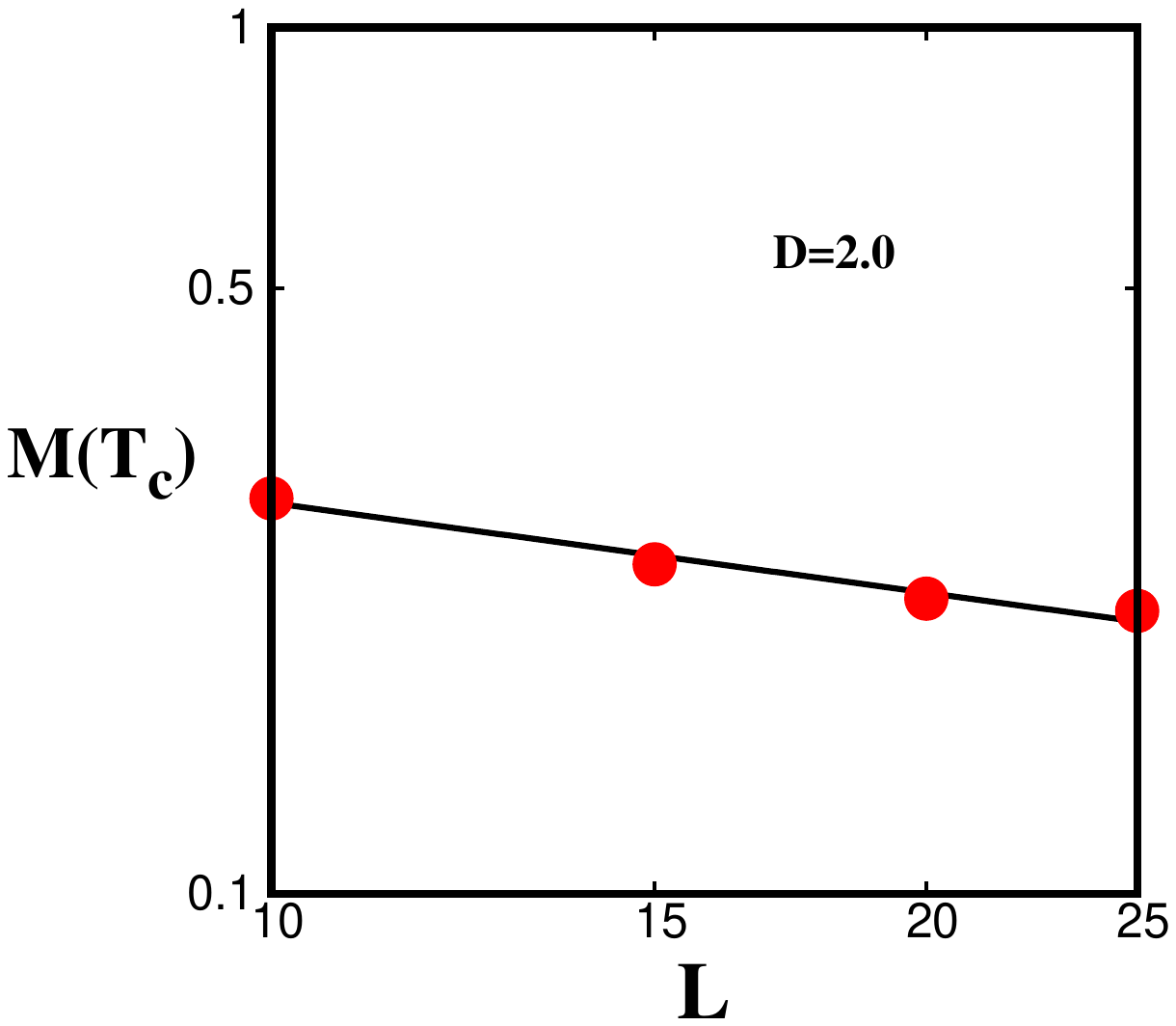}
    \caption{ \small{(a) Magnetisation ($M$) as a function of temperature ($T$) for different system sizes $L$. (b) The magnetisation at critical temperature ($M(T_c)$) as function of system sizes. The results are shown in log-log scale.}}
    \label{fig:const-D-M-L}
\end{figure}
The magnetic exponent ratio $\frac{\beta}{\nu}$ is determined from the scaling behaviour of magnetisation at the critical point via $M(T_c)\sim L^{-\frac{\beta}{\nu}}$. In Figure-\ref{fig:const-D-M-L}(b) we show the scaling behaviour in log-log scale. The critical exponent is obtained from the straight line fit as $\frac{\beta}{\nu}=0.344\pm0.04$.

\begin{figure}[h!]
    \centering
    (a)\includegraphics[width=0.45\linewidth]{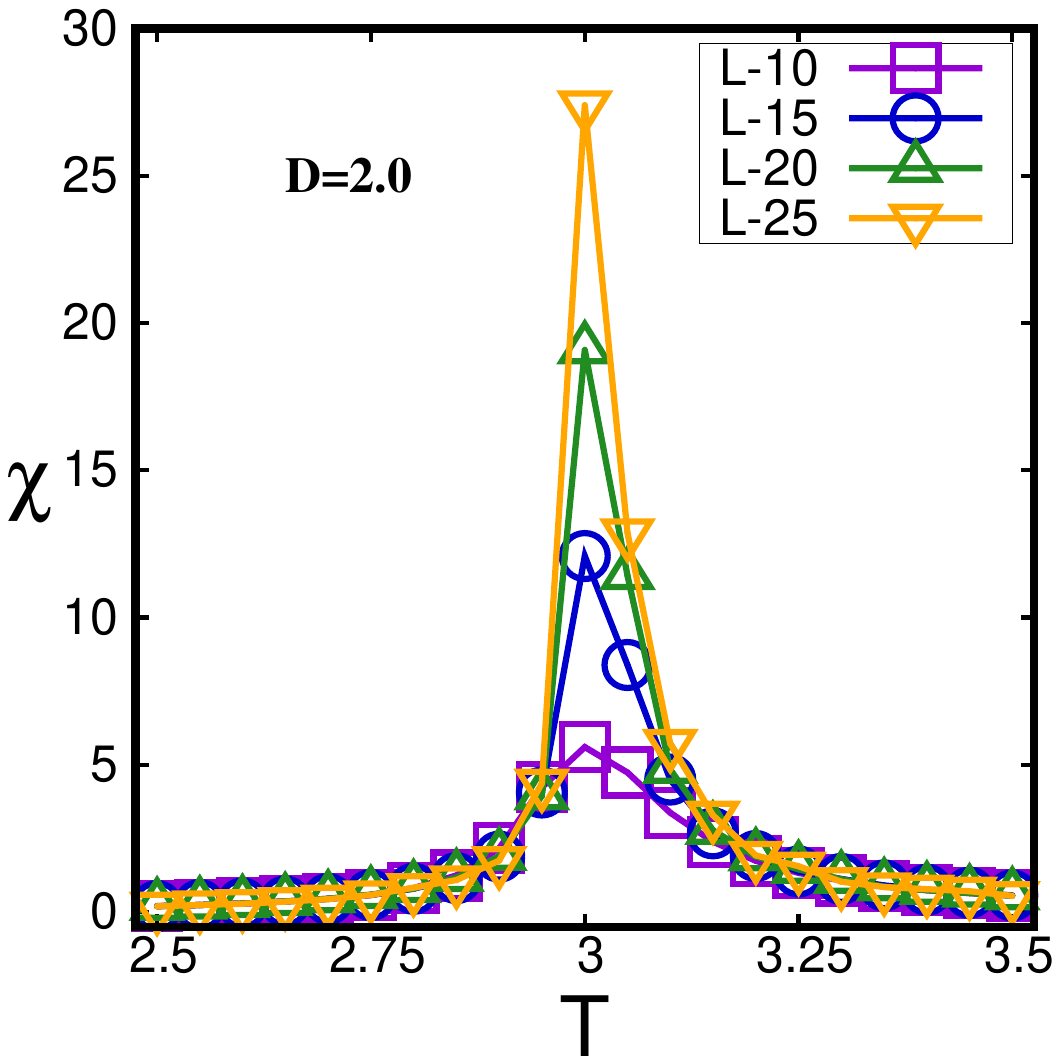}
    (b)\includegraphics[width=0.45\linewidth]{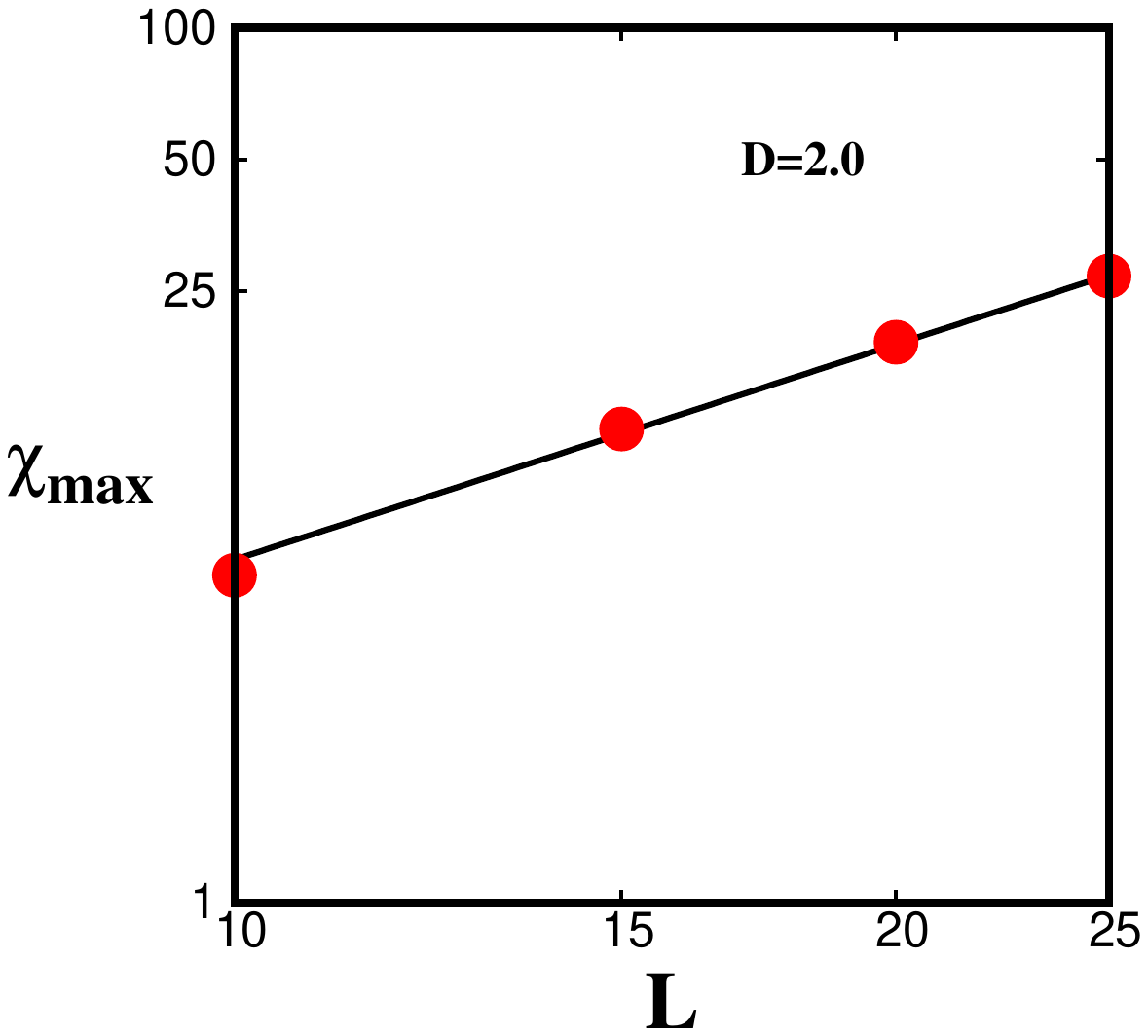}
    \caption{\small{ (a)Susceptibility ($\chi$) as a function of temperature ($T$) for different system sizes $L$. (b) The maxima of susceptibility ($\chi_{max}$) as function of system sizes. The results are shown in log-log scale.}}
    \label{fig:const-D-chi-L}
\end{figure}

The divergence of the susceptibility at the transition temperature is a crucial phenomenon in the equilibrium
continuous phase transitions. In addition to magnetisation, the corresponding susceptibility curves as a function of temperature are shown in Figure-\ref{fig:const-D-chi-L}(a). This figure illustrates the divergence of susceptibility with increasing system sizes, which validates the second-order phase transition. Figure-\ref{fig:const-D-chi-L}(b) shows the finite-size scaling behaviour of the maxima of susceptibility $\chi_{max}$. Assuming the scaling law as $\chi_{max} \sim L^\frac{\gamma}{\nu}$, $\chi_{max}$ is plotted against system size $L$ in log-log scale. The slope of the straight line fit estimates the critical exponent as $\frac{\gamma}{\nu}=1.636 \pm 0.057$. 

\vskip 1cm

\noindent{\large{\textbf{B.Quenched Random Field}}}\\
\vskip 0.2 cm
In this section critical properties of single-site anisotropic XY ferromagnet in presence of quenched random field have been presented. It is already shown in \cite{oliviarfxy} that random magnetic field plays a role of random disorder and leads to conventional ferro-para phase transition even in three-dimensional XY ferromagnet in presence of bilinear exchange anisotropy. Accordingly, we carried out MC simulation and studied XY ferromagnet in three dimension in presence of constant single-site anisotropy ($D$). We commence our analysis by presenting the dependency of the magnetisation and the corresponding susceptibility and specific heat on the random field strength (h) at  in Figure-\ref{fig:D-h-observable}. It is worth mentioning that $\langle h_i \rangle=0$, the randomness is incorporated by the direction of the field only. As we cool the system slowly from a high-temperature paramagnetic phase the magnetisation takes finite value continuously. The corresponding susceptibility shows a typical peak near the phase transition indicating the existence of the second-order phase transition in the system (\ref{fig:D-h-observable}(b)). The position of the peak of susceptibility determines the critical temperature of the phase transition. The specific heat curve also exhibits a peak at the vicinity of transition temperature (\ref{fig:D-h-observable}(c)).

\begin{figure}[]
    \centering
    (a)\includegraphics[width=0.55\linewidth]{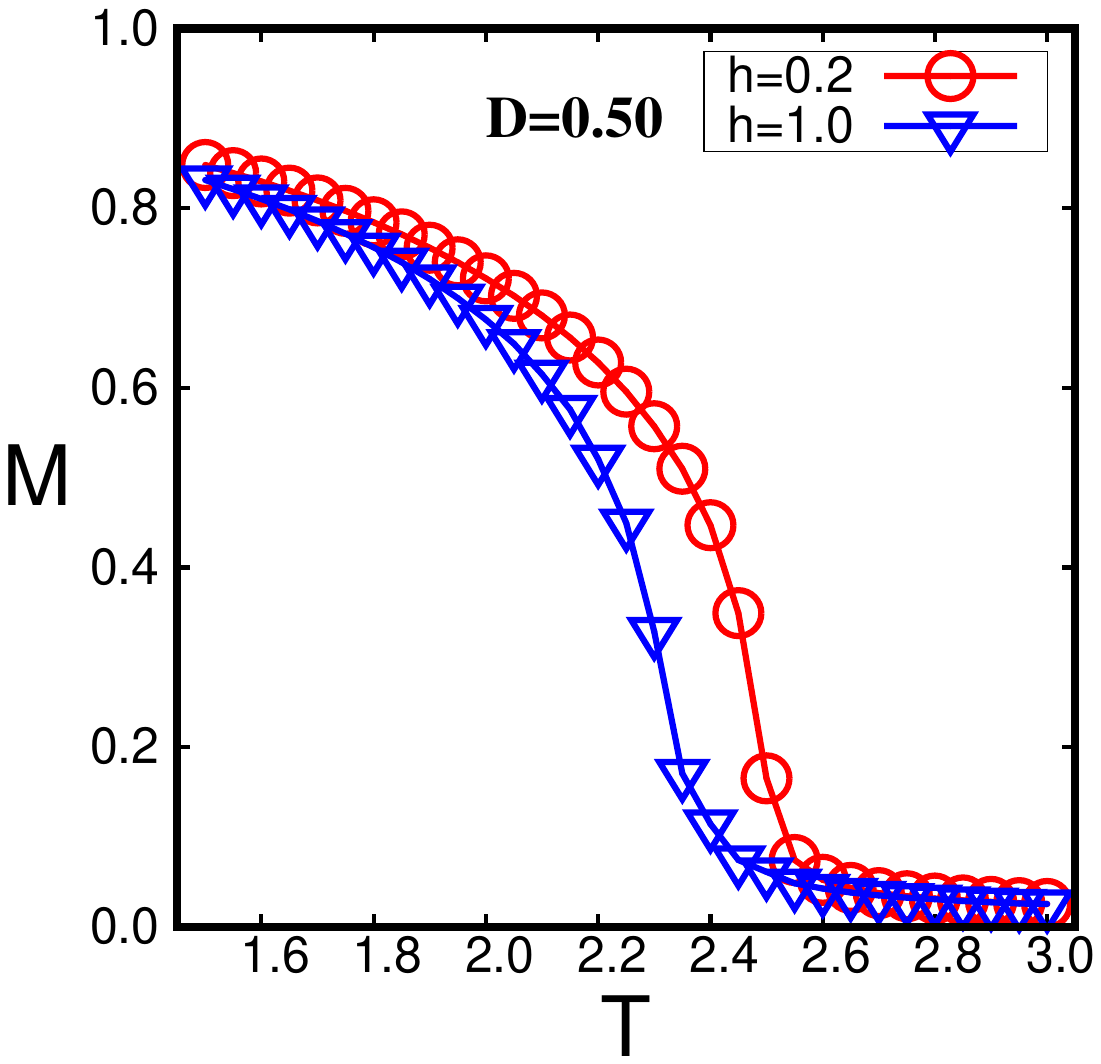}
    (b)\includegraphics[width=0.55\linewidth]{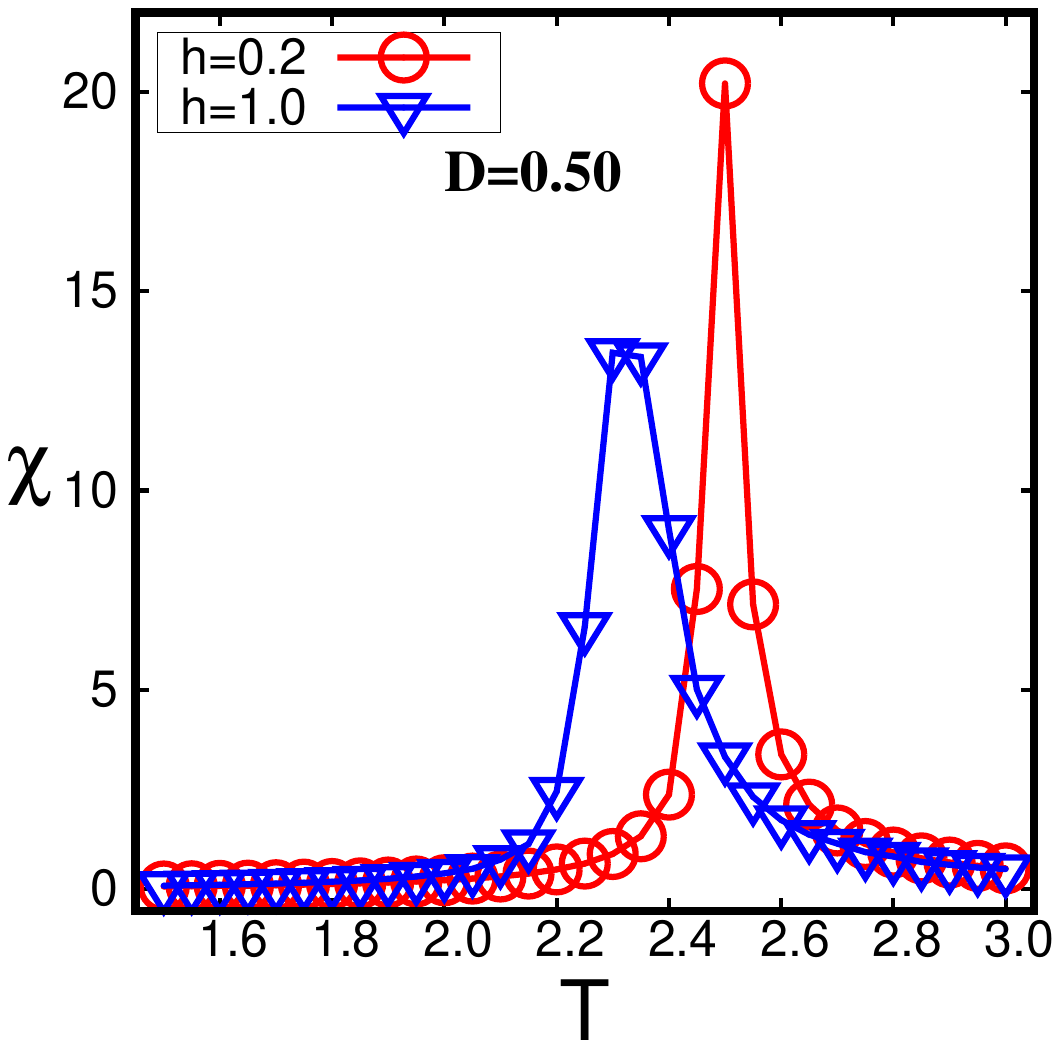}
    (c)\includegraphics[width=0.55\linewidth]{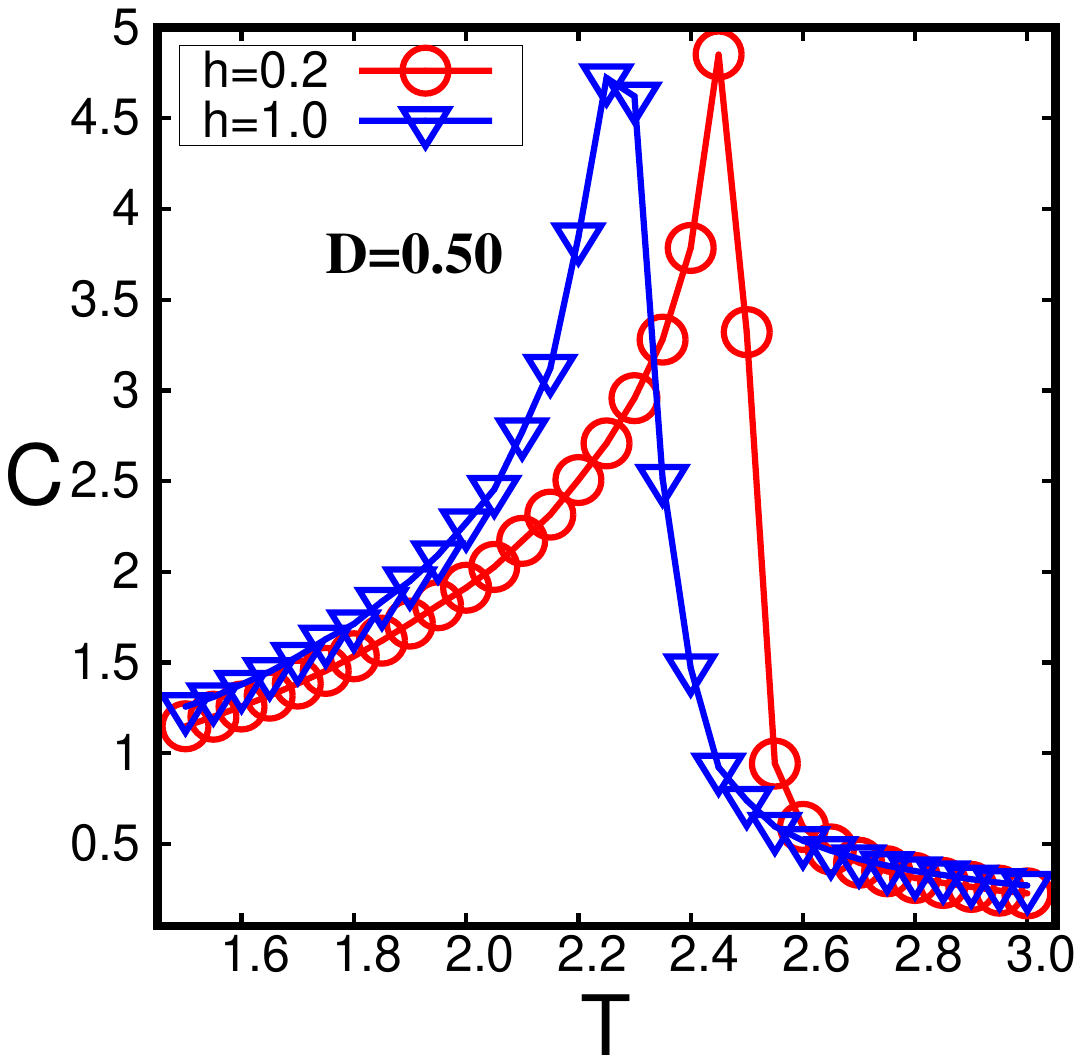}
    \caption{\small{ The thermal variation of (a) magnetization ($M$), (b) susceptibility ($\chi$) and (c) specific heat ($C$) for the system with $D = 0.50$ for different strength of random fields.}}
    \label{fig:D-h-observable}
\end{figure}

Thus single-site anisotropic XY ferromagnet also shows the ferro-para phase transition in presence of quenched random field even in three dimensions. From figure-\ref{fig:D-h-observable}(b) one can observe that for a particular value of anisotropy, as the strength of the random field is increased the transition points move to lower temperatures. This is due to the fact that the random field disrupts the alignment of spins in anisotropic XY ferromagnet by introducing disorder. The random field introduces local random fields that compete with the system’s tendency to align spins in a particular orientation. As a result the energy cost of aligning spins in presence of both constant anisotropy and random field increases, leading to a reduction in transition temperature.

\begin{figure}[ht!]
    \centering
    \includegraphics[width=0.65\linewidth]{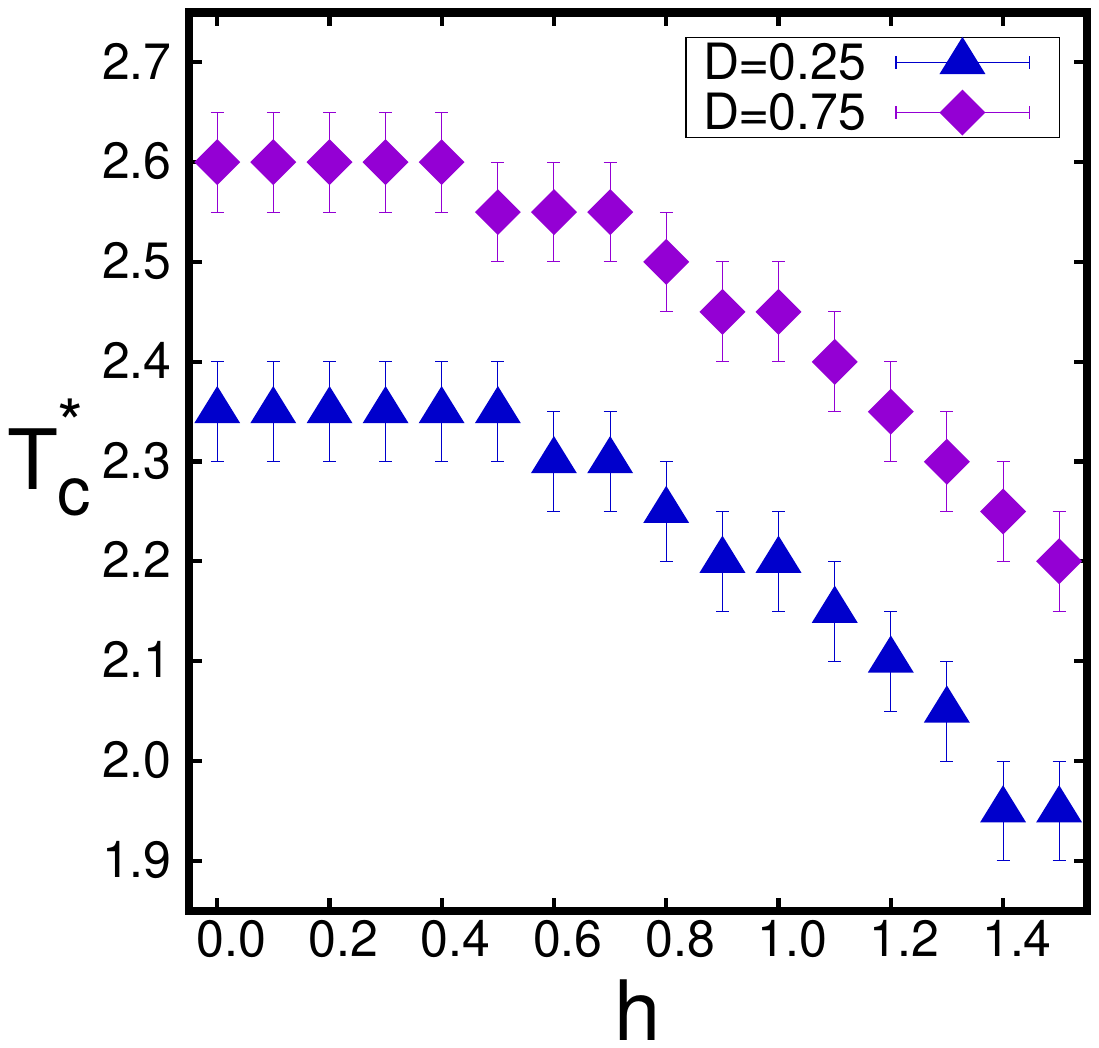}
    \caption{\small{The pseudo-critical temperature (${T_c}^*$) (obtained from the peak position of 
the susceptibility $\chi$) plotted as a function of random field strength ($h$) for different values of anisotropy ($D$)}.}
    \label{fig:Dtype-h-Tc}
\end{figure}
In previous section, we have shown that the constant single-site anisotropy increases critical temperature by reinforcing the alignment of spins in preferred direction, enhancing the stability of the ordered phase and require higher thermal energy to disrupt it. In contrast, the random field decreases the critical temperature by introducing a disorder to the system.Figure-\ref{fig:Dtype-h-Tc} depicts the pseudo-critical temperature as a function of random field strength for two different values of anisotropy. {\color{blue}It should be noted that, for small $h$, $T_c$ remains nearly constant, while for larger $h$, it decreases linearly. One may try to find the typical value of the random field required this crossover (from constant $T_c$ to linearly decreasing $T_c$). An interesting study may originate from this to determine the crossover field as a function of the anisotropy. }
\begin{figure}[ht!]
    \centering
    \includegraphics[width=0.65\linewidth]{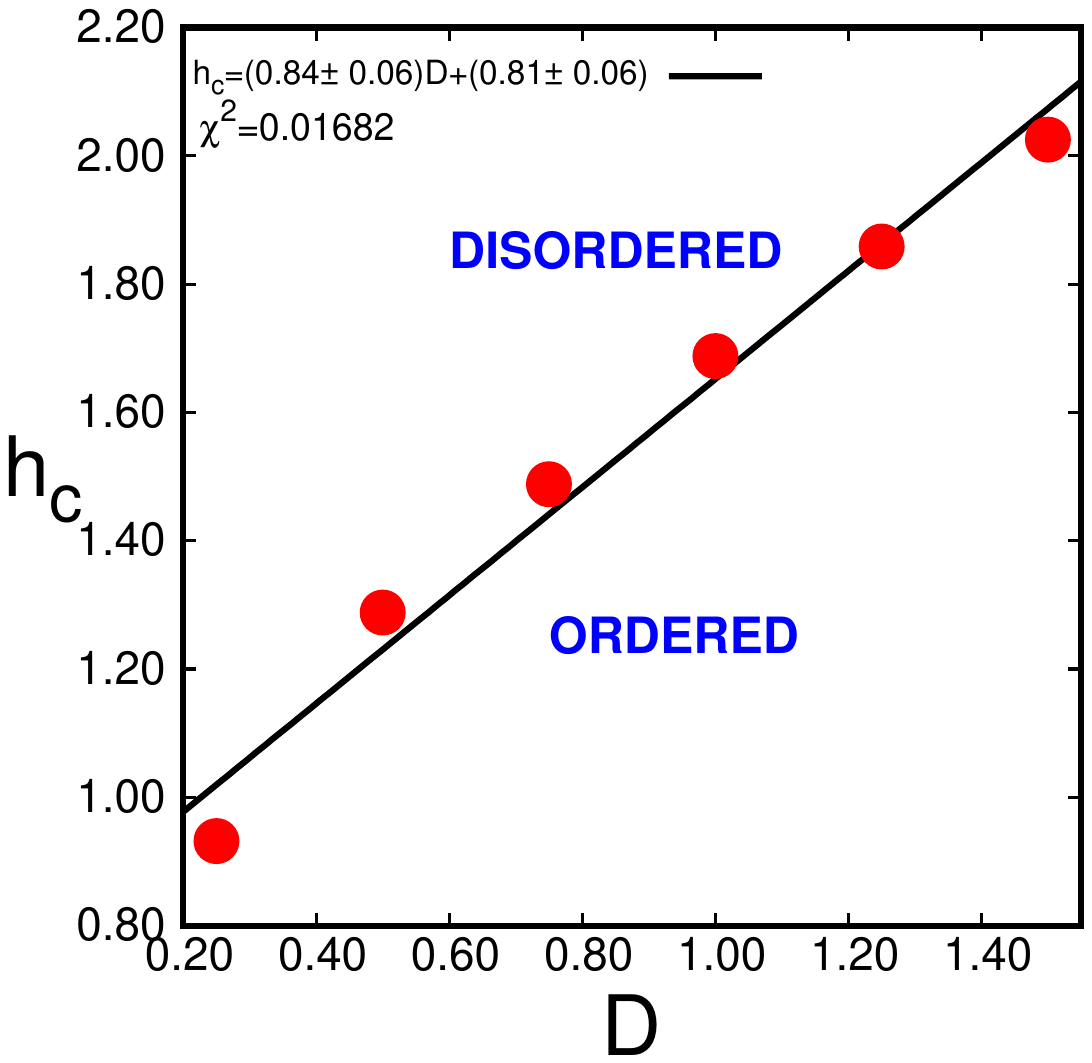}
    \caption{\small{The compensating field ($h_c$) is plotted as function of anisotropy strength ($D$). The compensating field has been calculated from the linear interpolation around the true critical temperature ($T_c = 2.206$) of the isotropic XY system. The data points are fitted with a straight line. The regions of ordered and disordered phases are separated by this line.}}
    \label{fig:Dtype-comfld}
\end{figure}

  {\color{blue}The constant anisotropy in the system introduces directional bias leading to an increase in the critical temperature($T_c$). In contrast, random field introduces disorder to the system, which reduces the transition temperature. So, here also a similar competition (as observed in the case of bilinear exchange anisotropy\cite{oliviarfxy}) between constant anisotropy and the quenched random field arises subject to the transition temperature.
In this context, we define compensating field ($h_c$): that is the amount of random field required to neutralize the effect of anisotropy, such that the system’s transition temperature compares with that of the isotropic XY system ($T_c$ = 2.206)\cite{campostrini}.
To determine the compensating field, we first compute the critical temperature for different values of random field strength and anisotropy. By linear interpolation, these values near the critical temperature of the isotropic system, I obtain the corresponding precise value of $h_c$.} The pairs of ($D- h_c$) are plotted in Figure-\ref{fig:Dtype-comfld}. The compensating field linearly depend on the strength of anisotropy ($D$). The region below the straight line corresponds to the ferromagnetically ordered phase and the above region is the disordered phase.

\vskip 1cm
\noindent {\bf {\Large 4. Discussion}}\\
\vskip0.2cm
While bilinear exchange anisotropy involves interactions between neighbouring spins, destroying the inherent SO(2) symmetry of the XY system, single-site anisotropy ($D$) similarly disrupts this SO(2) symmetry but accessories the extra energy contribution at individual spin sites. The motivation to study single-site anisotropy lies in its significant influence on the critical behaviour and phase transitions of magnetic materials, where local anisotropic effects play a pivotal role, akin to the impact seen with bilinear anisotropy. In this study, we have shown that the pseudocritical temperature increases linearly with strength of single-site anisotropy. This is because, the constant $D$ promotes bias towards ordering that requires more thermal energy to transit into the disordered phase, thereby the pseudocritical temperature increases. However, a contrasting scenario emerges when the anisotropy is randomly distributed in the system. The study examines random anisotropy with specified statistical distributions, including uniform, Gaussian, and bimodal distributions. Results show that pseudocritical temperature decreases as distribution width increases. Distributed anisotropy prevents bias toward ordering in spin-spin interactions, acting as a random disorder and reducing pseudocritical temperatures. 

Another form of disorder effect is reported in this work in form of quenched random field. This paper investigates the effects of quenched random fields on single-site anisotropic XY ferromagnets in three dimensions. Results showed in \cite{oliviarfxy} that there is a conventional ferro-para phase transition in presence of random fields, even in three dimensions with anisotropy. Anisotropy destroys the system's inherent SO(2) symmetry. The higher the random field magnitude, disorder-order transitions occur in lower temperatures. The compensating field ($h_c$) is defined to conserve the critical temperature for the \textit{isotropic} XY system. The compensating field increases linearly with anisotropy strength. The thermodynamic phase transition is formalized through finite size analysis, and the critical temperature is obtained in the thermodynamic limit. Critical exponent ratios are estimated.
\vskip 1.5cm

\noindent {\bf {\Large Acknowledgements:}}  I want to express my sincere gratitude to my supervisor Muktish Acharyya for his guidance, invaluable advice and continuous support during the present research work. I am thankful to MANF, UGC, Govt of India for financial support.
\vskip 0.5cm

\noindent {\bf Data availability statement:} Data will be available on request to Olivia Mallick.

\vskip 0.5cm

\noindent {\bf Conflict of interest statement:} I declare that this manuscript is free from any conflict of interest. The author has no financial or proprietary interests in any material discussed in this article.

\vskip 0.5cm

\noindent {\bf Funding statement:} No funding was received particularly to support this work.

\end{document}